\documentclass[12pt]{myelsart}
\usepackage{graphicx,subfigure,amsmath,amssymb,float}

\begin{document} 
\bibliographystyle{elsart-num}
\thispagestyle{empty}

\begin{frontmatter}
\title{Anisotropy-driven transition from collisionless to collisional regime
in the dipolar modes of a trapped gaseous mixture}

\author{P. Capuzzi\corauthref{cor1}},
\author{P. Vignolo},
and  \author{M.P. Tosi}
\corauth[cor1]{Corresponding author, e-mail: {\tt capuzzi@sns.it}}
\address{NEST-INFM and Scuola Normale Superiore,
Piazza dei Cavalieri 7, I-56126 Pisa, Italy}
\maketitle
\begin{abstract}
We evaluate the dipolar oscillations of a harmonically trapped 
fermion gas containing thermal bosonic impurities as a function 
of the anisotropy of the trap, from the numerical solution of the 
Vlasov-Landau equations for the one-body phase-space distribution functions. 
Starting from a situation in which the two components of the gaseous 
mixture perform almost independent oscillations inside a spherical trap, 
we demonstrate that different collision behaviors arise for oscillations 
in different directions as the trap is deformed into an elongated 
cigar-like shape. An increase in the anisotropy
of the confinement thus suffices to 
drive a transition of dipolar modes from a collisionless to a 
collisional regime.
\end{abstract}
\begin{keyword} 
Fermi gas, collisional properties, extended Kohn theorem
\vspace{0.5cm}
\end{keyword}
\thispagestyle{empty}
\end{frontmatter}

\clearpage

\section{Introduction}
Many aspects of the physics of ultracold quantum gases and their
quantum phase transitions can usefully be explored by measuring the
frequencies and the damping rates of monopolar "breathing" modes and
of quadrupolar modes at various temperatures (see Ref.
\cite{Minguzzi2004a} and references therein).  These measurements,
which have been performed with rather high accuracy on both bosonic
and fermionic systems, have been important to the understanding of
Bose-Einstein condensation and of degenerate and superfluid fermion
gases in confined geometries, and have become a testing ground for
many-body theories. 

In contrast, the dipolar "sloshing" modes in a monatomic gas under
harmonic confinement contain no deep information on its microscopic
behavior. These modes obey a theorem stemming from the work of Kohn
\cite{Kohn1961a} on a fluid of charged particles with arbitrary
momentum-conserving interactions in a magnetic field, which was
generalized by later workers \cite{Brey1989a} by adding an external
scalar harmonic potential extending throughout all space.  The theorem
states that in a system of interacting particles confined by an
external harmonic potential $V_h(\mathbf{r}) =
\mathbf{r}\cdot\mathbf{K} \cdot\mathbf{r}/2$ the dynamics of the
center of mass is completely decoupled from that of the internal
degrees of freedom. This implies sharp resonances at the bare trap
frequencies for a uniform exciting field, which can only rigidly drive
the density distribution of the system. The so-called extended Kohn
theorem (EKT) has had important consequences in the development of
time-dependent density functional theory for the dynamics of electrons
in confined geometries such as quantum wells and quantum dots
\cite{Dobson1994a}.

Turning, however, to a gaseous mixture, it is evident that its dipolar
modes provide a direct signature of its collisionality
\cite{Gensemer2001a,Toschi2003a}.  In a regime of low collision rates
the components of the mixture are independently oscillating, whereas
they oscillate together when their collisionality is sufficiently
high. This dynamical transition from a collisionless to a collisional
(hydrodynamic) behavior has been followed both experimentally and
numerically in two-component fermion mixtures
\cite{Gensemer2001a,Toschi2003a}.  The role played in this context by
mobile impurities inside a fermion gas under spherical confinement has
also been studied by numerical means \cite{Capuzzi2004b}. Interatomic
collisions also play an important role in the development and
understanding of techniques for the cooling of gases in essentially
harmonic confinements down to ultralow temperatures
\cite{Petrich1995a,Streckner2003a,Hadzibabic2003a}.
	
In the present work we examine the role of the anisotropy of the
harmonic confinement in this dynamical transition for a fermion-boson
mixture simulating a gas of fermionic $^{40}$K atoms which contains a
small concentration of thermal $^{87}$Rb atoms.  We numerically solve
the Vlasov-Landau equations for the evolution of the phase-space
distribution functions within a particle-in-cell approach (see
Refs. \cite{Capuzzi2004b,Toschi2004c} for the technical details). In
particular, our results demonstrate that in a cigar-shaped harmonic
confinement, rigid oscillations of the centre of mass of the whole
mixture in the "soft" axial direction can coexist with independent
oscillations of the two components in a "hard" radial direction. This
result can be considered as a further extension of the EKT. 
	
The paper is organized as follows. In Sec. \ref{sec:phys} we describe
the mixture under study, while in Secs. \ref{sec:coll} and
\ref{sec:coll2} we analyze its dynamics under time-dependent drives.
Finally, Sec. \ref{sec:concl} offers some concluding remarks.

\section{The physical system} 
\label{sec:phys}
The system is a mixture of spin-polarized $^{40}$K atoms and $^{87}$Rb
atoms inside a magnetic trap and the interspecies interaction is
modeled by a strongly repulsive contact potential characterized by an
{\it s}-wave scattering length $a=2\times10^3$ Bohr radii.  The mixture is at
low temperature and the number of $^{87}$Rb bosons is kept low in
order to avoid the formation of a condensate, which may in turn reduce
the collisionality of the system \cite{Timmermans1998a}. Moreover,
this allows us to neglect the boson-boson interactions.

Due to the difference in masses and in internal states, fermions and
bosons experience different confining potentials.  These are given by
\begin{equation}
V_{F,B}({\mathbf r}) = \frac{1}{2}\,m_{F,B}\,\omega^2_{F,B}
(r^2+\lambda^2\,z^2)\,,
\label{Eq:pote}
\end{equation}
where we have assumed the same anisotropy parameter $\lambda$ for the
two species. In Eq.\ (\ref{Eq:pote}) $m_F$ and $m_B$ are the atomic masses
of fermions and bosons and $\omega_F$ and $\omega_B$ are their trap frequency
along the radial direction. The effects of the anisotropy will be
examined by varying $\lambda$ from 1 to 0.01 while keeping fixed the
average $\bar{\omega}_{F,B} = \omega_{F,B}\lambda^{1/3}$ of the trap
frequencies.  This value will be taken as the geometric average of the
frequencies in the experiments carried out at LENS on the same
mixture \cite{Ferlaino2003a}. In this way we intend to explore the
dependence of the dynamics solely on the anisotropy of the traps
without changing other properties of the gas such as the mean
densities and the chemical potentials.  For definiteness we also set
the temperature of the sample at $T= 0.2 T_F$ with
$T_F=\hbar\bar{\omega}_F(6 N_F)^{1/3}/k_B$ being the Fermi temperature
and $N_F$ the number of fermions.


\section{Collisionality near equilibrium}
\label{sec:coll}
The number of collisions per unit time ({\it i.e.} the collision rate)
in the mixture at equilibrium is both classically and
quantum-mechanically independent of $\lambda$ for fixed
$\bar{\omega}_{F,B}$.  However, a slight mismatch of the trap centers
can lead to a strong decrease in the collisionality, which can be
understood to be the result of the decrease in the overlap of the two
density profiles. In the classical limit, the collisionality in a
mixture of $N_F$ fermions and $N_B$ bosons reads
\begin{equation}
\Gamma_{\text{cl}} =\beta
N_B\,N_F\,\frac{\sigma}{\pi^2}\frac{K_r^{3/2}}{m_r^{1/2}} \,
\exp\left[-\beta \frac{K_r}{2\,\lambda^{2/3}}(x_0^2 +
z_0^2\,\lambda^2)\right]
\label{eq:classcoll}
\end{equation}
when the trap centers have been offset by ($x_0$, $z_0$).
Here $\beta=1/k_B T$, $\sigma=4\pi\,a^2$ is the total scattering
cross section, $m_r$ the reduced boson-fermion mass,
$K_r= K_F K_B/(K_F+K_B)$ the effective oscillator
strength, and $K_{F,B} = m_{F,B}\,\bar{\omega}_{F,B}^2$ the average
oscillator strength of each species.  
Taking $x_0$ and $z_0$ to be of comparable magnitudes,
the collisionality as a function of the 
anisotropy thus displays two different trends depending on the 
direction of the offset.

\begin{figure}
\includegraphics[width=\columnwidth,clip=true]{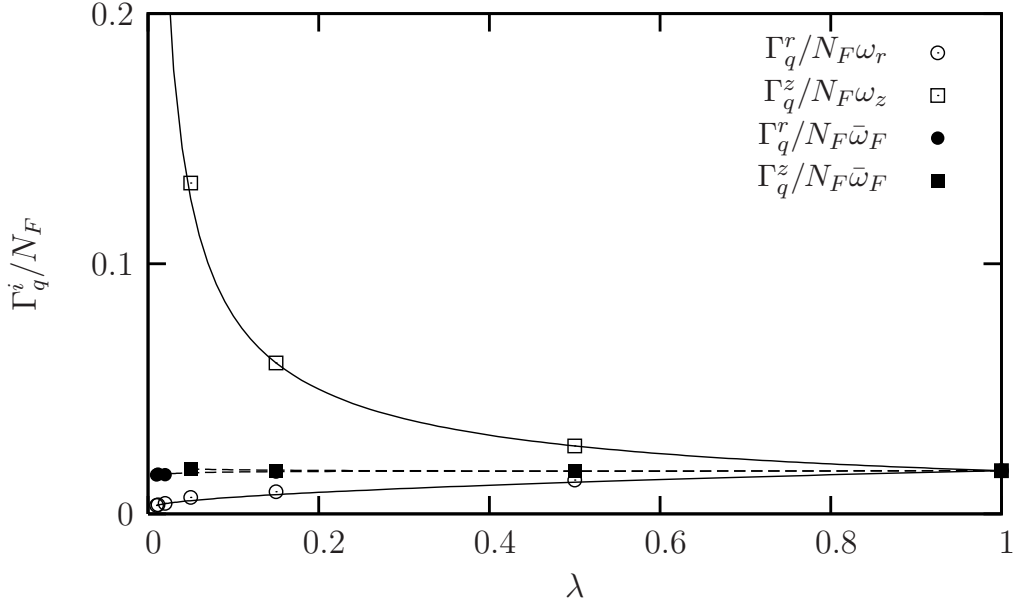}
\caption{\label{fig:Gammavslambda}  Total collision rates
$\Gamma_q^i$ per fermion in the radial ($i = r$) and axial ($i = z$) 
directions, as functions of the anisotropy parameter
$\lambda$. Filled symbols show the collision rates in units of
$\bar{\omega}_F$ and empty symbols report
$\Gamma_q^i/(N_F\omega_i)$. The squares refer to oscillations along the
radial direction and the circles to oscillations along the axis of the
trap. The lines are guides to the eyes.}
\end{figure}

In the numerical simulations we first shift rigidly the fermionic 
cloud in either the radial or the axial direction by 
$a_{ho}=(\hbar/m_F\omega_F)^{1/2}$, which is usually much smaller 
than the radius of the two clouds. Then the fermion cloud starts 
oscillating around its equilibrium position and 
collisions among bosons and fermions can occur.
During the evolution the relative position of 
the centers of the two clouds changes continuously and
Eq.\ (\ref{eq:classcoll}) gives only an indication of the dependence
of the time-averaged collisionality on $\lambda$. 
From the simulation data we have calculated 
the average number of collisions per unit time during a fixed 
time interval, that we choose to be $\bar{t}=10\pi/\bar{\omega}_F$.
The result is shown in Fig. \ref{fig:Gammavslambda},
where we plot the average quantum collision rates   
$\Gamma_q^z$ and $\Gamma_q^r$, corresponding to dipolar 
oscillations  of the fermions in the two directions, as functions 
of the trap anisotropy $\lambda$.
The filled symbols
refer to the collision rates in units of the constant frequency scale
$\bar{\omega}_F$. 

The data in Fig. \ref{fig:Gammavslambda} show that for the system 
parameters that we are considering the variation 
with $\lambda$ of the two collision rates $\Gamma_q^i$ 
in these absolute units becomes appreciable only at low $\lambda$. 
However, referring to the empty symbols in Fig. \ref{fig:Gammavslambda} 
which report $\Gamma_q^z$ and $\Gamma_q^r$ in units of the 
fermionic axial and radial frequencies 
$\omega_z=\lambda^{2/3}\bar{\omega}_F$ and
$\omega_r=\lambda^{-1/3}\bar{\omega}_F$, it is seen that the numbers 
of collisions made on average during an oscillation period in the 
axial or in the radial direction are becoming very different.
This suggests that, since $\Gamma_q^r/\omega_r \ll 1$ while
$\Gamma_q^z/\omega_z$ 
is strongly increasing, the mixture may have a collisionless 
behavior along the radius $r$ and a hydrodynamic 
behavior along the $z$ axis.

\begin{figure}
\includegraphics[width=\columnwidth,clip=true]{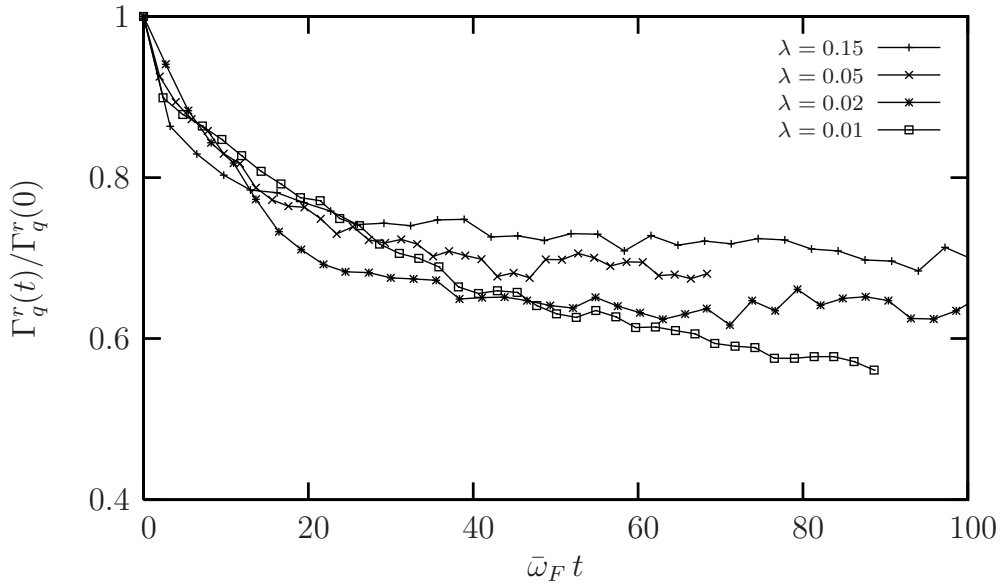}
\caption{\label{fig:Gammavst} Total collision rate
$\Gamma_q^r(t)/\Gamma_q^r(0)$ as a function of time for several values
of the anisotropy $\lambda$ and in the case of an initial radial
displacement.}
\end{figure}

The return of the system to a new equilibrium state 
(including instantaneous mean-field effects) can be directly 
investigated in the Vlasov-Landau frame by following the collision 
rate as a function of time. As the evolution proceeds the 
average number of collisions approaches that 
corresponding to the new equilibrium. We display in Fig. \ref{fig:Gammavst} 
the time dependence of the radial collision rate for different 
values of $\lambda$. The collision rate 
diminishes relatively rapidly towards the final equilibrium state, as 
can be expected from the temperature dependence of $\Gamma_{\text{cl}}$  
in Eq. (\ref{eq:classcoll}) since the system performs a number-conserving 
evolution.  

\section{Collisionless {\em vs} collisional behaviour} 
\label{sec:coll2}
Let us summarize some general features of the dynamical behavior of 
the mixture that are associated with an initial displacement of the 
fermion cloud, before turning to discuss the role of 
trap anisotropy. In the absence of interactions the fermions oscillate 
without damping at their bare trap frequency while the bosonic 
impurities remain at rest. On switching on the interactions 
the fermionic components sets the bosons into motion and beats 
between oscillations at the 
bosonic and the fermionic trap frequencies appear~\cite{Toschi2004c}.
If the collision rate is much lower than 
the trap frequencies, the damping rate is small and the 
two clouds oscillate for many periods. On 
increasing the collisionality up to a regime where the collision rate 
is of the same order as the trap frequencies, the oscillation 
frequencies are not well defined and the center-of-mass motions 
of the two clouds become strongly damped. Finally, at very large 
collisionality the two clouds become glued together and oscillate 
at the same frequency without noticeable damping. Thus a 
clear signature of the onset of a collisional regime is the locking 
of the oscillations of the two components and the vanishing of 
their damping rates.

\begin{figure}
\hspace{-1.5cm}\subfigure[]{\includegraphics[width=0.6\linewidth,clip=true]{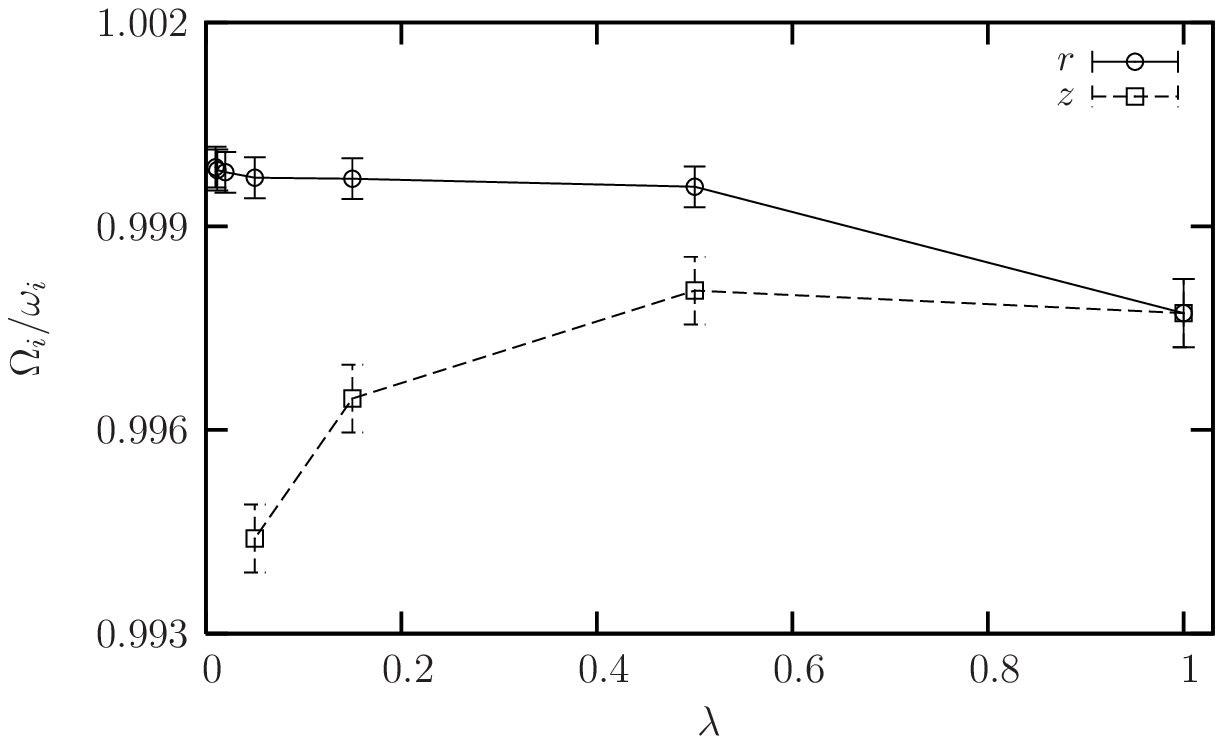}}%
\subfigure[]{\includegraphics[width=0.6\linewidth,clip=true]{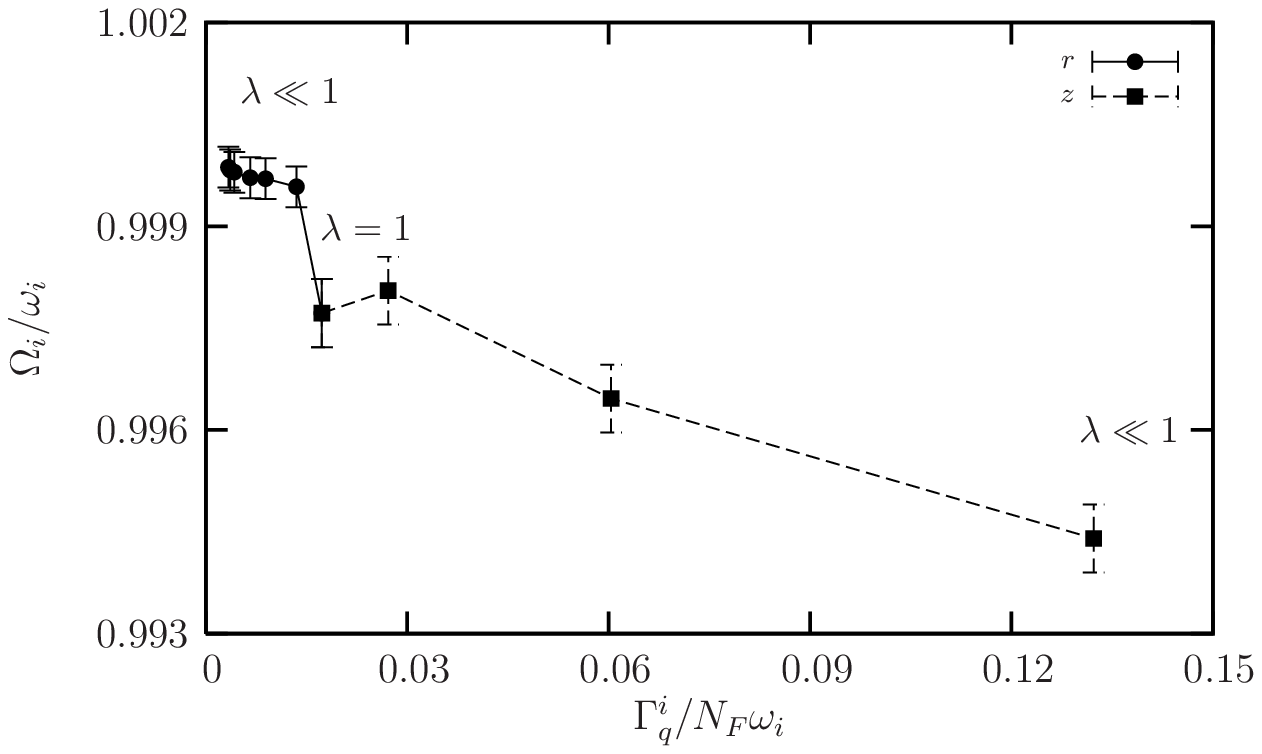}}
\caption{ Radial and axial oscillation frequencies (in units of
$\omega_r$ and $\omega_z$, respectively) as functions of (a) the
anisotropy parameter $\lambda$, and (b) the collision rate
$\Gamma_q^i/N_F$ per fermion (in units of $\omega_r$ for radial
oscillations and of $\omega_z$ for axial oscillations).}
\label{fig_osc}
\end{figure}

We turn now to examining the role played by trap anisotropy in this scenario.
We plot in Fig. \ref{fig_osc}(a) the oscillation frequencies of the fermions in the radial and axial direction as functions 
of the anisotropy parameter $\lambda$ which drives the collision rates 
referred to the trap frequencies (see Fig. \ \ref{fig:Gammavslambda}).
The data in Fig. \ref{fig_osc} have been obtained by fitting the oscillatory motions of the 
fermions with functions of the form $\cos(\Omega\, t + \phi)\exp(-\gamma\,t)$. 
The oscillation frequency $\Omega_r$ in the radial direction
approaches the radial trap frequency $\omega_r$ for low $\lambda$ and 
decreases below this value with increasing $\lambda$. On the contrary, the 
oscillation frequency $\Omega_z$ in the axial direction takes its 
maximum value at $\lambda=1$ and decreases with decreasing $\lambda$.
These two opposite trends can be understood if we plot the oscillation 
frequencies as functions of the collision rates scaled by the 
trap frequencies. In such a plot (see Fig. \ref{fig_osc}(b))
both oscillation frequencies lie on a single 
monotonically decreasing  curve. The plot suggests that the leftmost 
data correspond to collisionless radial oscillations while the rightmost 
data reflect an approach to the collisional regime where the axial 
oscillation frequencies are close to the hydrodynamic value.

\begin{figure}
\centerline{\includegraphics[width=0.8\linewidth,clip=true]{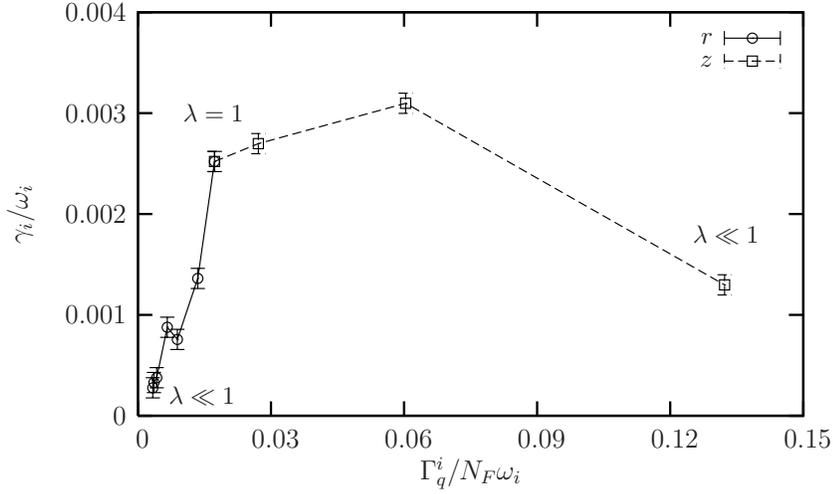}}
\caption{Radial (circles) and axial (squares) damping rates $\gamma_i$
(in units of $\omega_r$ and $\omega_z$, respectively) as functions of
the collision rates $\Gamma_q^r/(N_F\omega_r)$ and $\Gamma_q^z/(N_F\omega_z)$.}
\label{fig_damp}
\end{figure}

To support the idea that the mixture at large trap anisotropies is in
different regimes in the radial and in the axial direction, we report
in Fig. \ref{fig_damp} the damping rates $\gamma_r$ and $\gamma_z$ for
the oscillatory motions as functions of the corresponding collision
rates. In the radial direction (circles in Fig. \ref{fig_damp}) the
damping rate increases as a function of the scaled collision rate
$\Gamma_q^r/\omega_r$ and seems to reach a maximum: this is a
signature that the gas is moving from the collisionless to the
intermediate collisional regime with decreasing anisotropy. On the
contrary, in the axial direction (squares in Fig. \ref{fig_damp}) the
damping rate decreases as a function of $\Gamma_q^z/\omega_z$ and this
is a typical behavior of an intermediate-to-collisional transition.

\section{Conclusions}
\label{sec:concl}
We have studied the dipolar modes of a harmonically confined fermion-boson 
mixture as functions of the trap anisotropy and found that, although the 
collisionality is independent of the anisotropy, a collisionless behavior 
in the radial direction and a hydrodynamic behavior in the axial direction 
can be simultaneously established for large anisotropies. 
This conclusion is based on the analysis of the dipolar oscillations of 
a fermion gas at ultralow temperature and containing a limited 
concentration of thermal bosons.
	
The analysis that we have presented has only regarded the dipolar
oscillations of the gas. Of course, other oscillatory modes and the
dynamics of a ballistic expansion would provide further insight on the
physical behavior of the gas. The results of these further studies
will be presented elsewhere.

\ack 
This work has been partially supported by an Advanced Research Initiative 
of Scuola Normale Superiore di Pisa and by the Istituto Nazionale di 
Fisica della Materia within the Advanced Research Project "Photonmatter" 
and the Initiative "Calcolo Parallelo".


\end{document}